\documentstyle[sprocl]{article}                                                

\input{psfig}
\def\Journal#1#2#3#4{{#1} {\bf #2}, #3 (#4)}

\def\NPB{{\em Nucl. Phys.} B}
\def\PLB{{\em Phys. Lett.}  B}
\def\PRD{{\em Phys. Rev.} D}
\def\ZPC{{\em Z. Phys.} C}
\def\be{\begin{equation}}
\def\ee{\end{equation}}
\def\bea{\begin{eqnarray}}
\def\eea{\end{eqnarray}}

\begin{document}

\title{HARD AND SOFT COLOUR SINGLET EXCHANGE IN THE SEMICLASSICAL APPROACH}

\author{A. HEBECKER}

\address{D.A.M.T.P., Cambridge University, Cambridge CB3 9EW, England}

\maketitle\abstracts{In the present talk diffraction in deep inelastic 
scattering is discussed in the framework of the semiclassical approach. 
The main emphasis is on the possibility of a consistent semiclassical 
description of both hard and soft colour singlet exchange processes. This 
approach allows the comparison of hard and soft colour neutralization in 
diffractive electroproduction of high-$p_\perp$ jets or heavy quarks.}

\section{Introduction}
One of the most interesting theoretical problems of diffraction in deep 
inelastic scattering is the precise nature of the $t$-channel colour 
singlet exchange, which is the main characteristic of this type of 
processes. On the one hand, the presence of the hard scale $Q^2$, provided 
by the virtual photon, and possibly additional hard scales, such as 
high-$p_\perp$ jets or heavy quarks in the final state, suggest the 
applicability of perturbation theory. From this point of view the natural 
mechanism would be perturbative two-gluon exchange in the $t$-channel 
\nolinebreak \cite{gtg}. On the other hand, it is well known that at high 
energy the wave function of a highly virtual photon develops a soft 
component which has a large, hadronic cross section and contributes to 
inclusive deep inelastic scattering at leading twist \cite{bk}. This clearly 
suggests a corresponding leading twist contribution to diffraction in which 
the hard scales from photon and diffractive final state do not render the 
colour singlet exchange perturbative. 

The above two mechanisms for colour singlet exchange result in different 
dependences of cross sections on $Q^2$, $x$ and possible additional 
parameters like $p_\perp$ or heavy quark masses in the diffractive final 
state. However, as will be discussed below, the relative normalization of 
both contributions remains unknown. 

In the present talk, which is based on \cite{bdhc,bdhp}, it will be shown 
how a unified description of both hard and soft colour singlet exchange 
naturally arises within the semiclassical framework. This observation can be 
taken as a starting point for the phenomenological analysis of 
high-$p_\perp$ jets and charm in diffraction \cite{mcd}. A more detailed 
discussion of the semiclassical approach to the above two diffractive 
processes and the resulting phenomenological predictions can be found in 
\cite{bdhc,bdhp}.

\section{Hard colour singlet exchange and squared gluon density}
\begin{figure}[ht]
\begin{center}
\parbox[b]{8cm}{\psfig{width=8cm,file=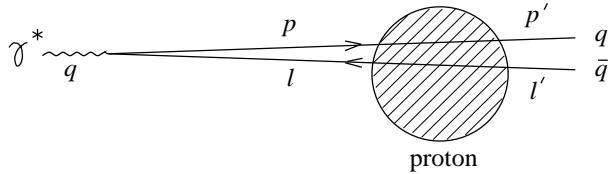}}\\
\end{center}
\caption{Production of a $q\bar{q}$-pair in the semiclassical approach.
\label{qq}}
\end{figure}

Before focusing on the description of hard colour singlet exchange recall 
the general picture of leading order diffraction in the semiclassical 
approach. Working in the proton rest frame and modelling the proton by a 
classical colour field the simplest process is the creation of a colour 
neutral $q\bar{q}$-pair by the virtual photon (see Fig.~\ref{qq}). For 
transverse photon polarization and for one massless quark generation with 
one unit of electric charge the corresponding cross section reads 
\cite{bdhp,bdh}
\bea
\left.{d\sigma_T\over dt}\right|_{t\approx 0}&=&\frac{\alpha_{em}}
{6(2\pi)^6}\int d\alpha dp'^2_{\perp} (\alpha^2 + (1-\alpha)^2) \times
\label{dsdt}
\\ \nonumber \\
&&\left| \int_{x_{\perp},y_{\perp},p_{\perp}}e^{iy_{\perp}
(p_{\perp}-p'_{\perp})}\mbox{tr}W^{\cal F}_{x_{\perp}}(y_{\perp})
\frac{p_{\perp}}{\alpha (1-\alpha)Q^2 + p_{\perp}^2}\right|^2,\nonumber
\eea
where $t=(q-p'-l')^2$ is the momentum transfer to the proton and 
$\alpha=p'_0/q_0$. 

The proton colour field is described by the quantity 
\be
W^{\cal F}_{x_\perp}(y_\perp)=U^\dagger(x_\perp+y_\perp)U(x_\perp)-1
\label{wa}
\ee
which is built from the non-Abelian eikonal factors $U$ and $U^\dagger$ of 
the quark and antiquark whose light-like paths penetrate the colour field of 
the proton at transverse positions $x_\perp$ and $x_\perp+y_\perp$, 
respectively. The superscript ${\cal F}$  is used because the quarks are in 
the fundamental representation of the gauge group. 

In this language the hardness of the process is characterized by the 
relevant transverse distances $|y_\perp|$. Large $Q^2$ in itself is not 
sufficient to make the colour singlet exchange hard since it can be 
compensated by small $\alpha$ or $1-\alpha$. However, the requirement of 
large $p_\perp'$ leads to the dominance of large $p_\perp$ and, eventually, 
to the dominance of small distances in the $y_\perp$-integration in 
Eq.~(\ref{dsdt}). The same effect can also be achieved by giving the quarks 
a large mass \cite{bdhc}, in which case the process remains hard for all 
$p_\perp'$. As a result, the leading order cross section is only sensitive 
to the first term in the Taylor expansion of tr$W^{\cal F}$, 
\be
\int_{x_{\perp}}\mbox{tr}W^{\cal F}_{x_{\perp}}(y_{\perp})
=\mbox{const.}\times y_{\perp}^2 + {\cal O}(y_{\perp}^4)\, .\label{tew}
\ee

\begin{figure}[ht]
\begin{center}
\parbox[b]{9.3cm}{\psfig{width=9.3cm,file=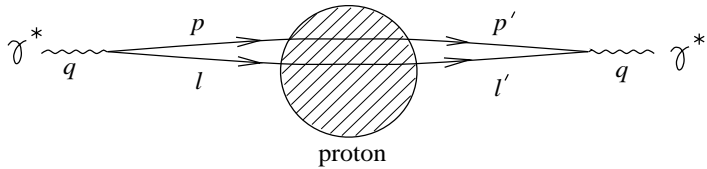}}\\
\end{center}
\caption{The Compton scattering amplitude within the semiclassical 
approach.
\label{comp}}
\end{figure}

In the leading-log approximation the coefficient of the above 
$y_\perp^2$-term can be related to the inclusive gluon density. To see this
observe that, within the semiclassical approach, the colour dipole cross 
section $\sigma(\rho)$ is given by \cite{dy}
\be
\sigma(\rho)=-\frac{2}{3}\int_{x_{\perp}}\mbox{tr}W^{\cal F}_{x_{\perp}}
(\rho_{\perp})\,.
\ee
Using the well-known relation of the short distance behaviour of 
$\sigma(\rho)$ and the gluon density \cite{fms},
\be
\sigma(\rho)=\frac{\pi^2}{3}\alpha_s[xg(x)]\rho^2+{\cal O}
(\rho^4)\,,
\ee
the constant in Eq.~(\ref{tew}) and hence the high-$p_\perp$ cross section 
of Eq.~(\ref{dsdt}) can be determined. This normalization of the 
$y_\perp^2$-term of tr$W^{\cal F}$ by the gluon density can also be derived 
by calculating the Compton amplitude of Fig.~\ref{comp} and relating it to 
inclusive deep inelastic scattering via the optical theorem. 

Combining the above formulae the following expression for the diffractive 
production of two high-$p_\perp$ jets can be derived
\be
\left.\frac{d\sigma_T}{dtd\alpha dp'^2_{\perp}}\right|_{t\approx 0}=
\frac{2\pi^2\alpha_{em}\alpha_s^2[xg(x)]^2}{3}\,\frac{[\alpha^2+
(1-\alpha)^2]\,[\alpha(1-\alpha)]^2Q^4p_\perp'^2}{[\alpha(1-\alpha)Q^2+
p_\perp'^2]^6}\,.\label{tjcs}
\ee
The identification of the squared gluon density in hard diffractive 
processes has been demonstrated in \cite{ryskin} in the framework of vector 
meson production. As expected, Eq.~(\ref{tjcs}) is in agreement with the 
corresponding two-gluon exchange calculations for diffractive jet production 
(see e.g. \cite{tg}).

\section{Soft colour singlet exchange and diffractive gluon density}
\begin{figure}[ht]
\begin{center}
\parbox[b]{8cm}{\psfig{width=8cm,file=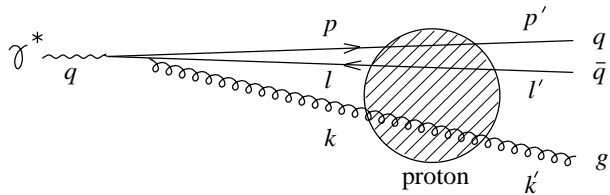}}\\
\end{center}
\caption{Diffractive jet production with an additional soft gluon in the 
final state.
\label{qqg}}
\end{figure}

The physics of the colour neutralization changes radically if, in addition 
to the two high-$p_\perp$ quark jets, a gluon is present in the 
diffractive final state. As has been shown in \cite{bdh} now the dominant 
contribution comes from the phase space region where this gluon is 
relatively soft, i.e.~it has small transverse momentum and carries only 
a small fraction of the longitudinal momentum of the photon. This means 
that the gluon develops a large transverse separation from the 
$q\bar{q}$-pair, thus testing the proton field non-perturbatively (see 
Fig.~\ref{qqg}). 

The process can be reinterpreted as high-$p_\perp$ jet production in 
boson-gluon fusion \cite{h} and the cross section
\be
\frac{d\sigma_T}{d\xi dp_\perp'^2}=\int_x^\xi dy
\frac{d\hat{\sigma}_T^{\gamma^* g\rightarrow q\bar{q}}(y,p_{\perp}')}
{dp_\perp'^2}\frac{dg(y,\xi)}{d\xi}\label{bgfu}
\ee
involves a diffractive gluon density 
\be
\frac{dg(y,\xi)}{d\xi}\!=\!\frac{1}{8\xi^2}\left(\!\frac{b}{1\!-\!b}\!
\right)\!\int\!\frac{d^2k'_\perp(k'^2_\perp)^2}{(2\pi)^4}\!\int_{x_\perp}
\left|\int\frac{d^2k_\perp}{(2\pi)^2}\frac{\mbox{tr}[
\tilde{W}^{\cal A}_{x_\perp}(k'_\perp\!-\!k_\perp)]t^{ij}}{k_\perp'^2 b+
k_\perp^2(1\!-\!b)}\right|^2\!\!,\label{gd}
\ee
\be
t^{ij}=\delta^{ij}+\frac{2k^i_{\perp}k^j_\perp}{k'^2_\perp}
\left(\frac{1-b}{b}\right)\, .
\ee
Here $\xi=x (Q^2+M^2)/Q^2$ for a final state with diffractive mass $M$, the 
momentum fraction of the proton carried by the incoming gluon is denoted by 
$y$, and $b=y/\xi$. The function $\tilde{W}^{\cal A}_{x_\perp}$ is the 
Fourier transform of $W^{\cal A}_{x_\perp}$ which is defined as in 
Eq.~(\ref{wa}) but with the $U$-matrices in the adjoint representation. 
Since Eq.~(\ref{gd}) does not involve the hard scales of the process the 
function $W^{\cal A}_{x_\perp}(y_\perp)$ is tested in the whole range of 
$y_\perp$. This is in contrast to Eq.~(\ref{dsdt}) where only the 
perturbative small-$y_\perp$ region matters in the high-$p_\perp$ limit. 

For a direct comparison with the exclusive two-jet production of 
Eq.~(\ref{dsdt}) the differential cross section, Eq.~(\ref{bgfu}), has to be 
rewritten in terms of $d\alpha$. This is easily done introducing the 
invariant mass of the two quark jet system, $M_j^2=p_\perp'^2/\alpha(1\!-\!
\alpha)$, and using the relation $yQ^2=x(Q^2+M_j^2)$. For more 
phenomenological details see \cite{mcd}.

\section{The relative normalization of hard and soft contribution}
So far it has been discussed how high-$p_\perp$ jets in diffractive 
electroproduction can be generated by either hard or soft colour singlet 
exchange mechanisms. The relative normalization of these two mechanisms is 
not easily predicted from first principles and should be determined from 
experiment. 

Consider first the case of hard colour singlet exchange described by 
Eq.~(\ref{dsdt}) and, more explicitly, by Eq.~(\ref{tjcs}). Although the 
cross section at $t\approx 0$ (or, more precisely, at the very small value 
$-t=m_p^2\xi^2$) can be expressed in terms of the squared gluon density this 
is not the case for the full cross section of diffractive jet production. 
The reason is that the $t$-integration introduces the unknown constant 
\be
C=\left(\int\frac{d\sigma}{dt}dt\right)\bigg/\left(\frac{d\sigma}{dt}
\bigg|_{t\approx0}\right)\sim \Lambda^2\, ,
\ee
where $\Lambda$ is a typical hadronic scale. 

The normalization of the contribution with soft colour neutralization is 
also not calculable. As can be seen from Eqs.~(\ref{bgfu}) and (\ref{gd}) 
the production of the hard jets is governed by the standard cross section 
for boson-gluon fusion. In contrast, the radiation of the soft gluon and 
the overall normalization is described by the non-perturbative quantity 
$dg(y,\xi)/d\xi$.

In spite of the fact that the normalization of the cross section is not 
known in both of the above cases some more general statements about the 
ratio of soft and hard contributions can be made. First of all, it is well 
known that the contribution with additional gluon is dominant in the region 
of large diffractive masses. It is, however, also interesting to consider 
the region of medium $M^2$ in more detail. 

Setting $M^2=Q^2$ (which corresponds to $\beta=1/2$) and integrating over 
all $p_\perp'^2>p'^2_{\perp,\mbox{\footnotesize min}}$ the following 
dependences on $Q^2$ and $p'^2_{\perp,\mbox{\footnotesize min}}$ are 
obtained for the exclusive two-jet and the two-jet plus gluon contribution 
respectively,  
\be
\left(\frac{d\sigma}{d\xi}\right)_{q\bar{q}}\sim\frac{1}{Q^2}\,
\frac{\Lambda^2}{p'^2_{\perp,\mbox{\footnotesize min}}}\,,\qquad
\left(\frac{d\sigma}{d\xi}\right)_{q\bar{q}g}\sim\frac{1}{Q^2}\,
\ln Q^2/p'^2_{\perp,\mbox{\footnotesize min}}\,.\label{te}
\ee
These rough estimates neglect additional dependences on the parameters 
introduced via the scales of factors of $\alpha_s$. Nevertheless, 
Eq.~(\ref{te}) clearly shows that for sufficiently large 
$p'^2_{\perp,\mbox{\footnotesize min}}$ the $q\bar{q}g$ final state, and 
hence the soft colour neutralization mechanism, is dominant even in the 
region of medium $M^2$.

\section{Conclusions}
In diffractive electroproduction of high-$p_\perp$ jets contributions from 
pure $q\bar{q}$ and $q\bar{q}g$ final states rely on different mechanisms 
of colour neutralization. While in the first case the colour singlet 
exchange is hard, corresponding to perturbative two-gluon exchange, in the 
second case the soft gluon implies a non-perturbative mechanism of 
colour neutralization. The semiclassical approach to diffraction provides a 
consistent framework for the treatment of both contributions. It can 
therefore serve as a starting point for the phenomenological analysis of 
the experimental cross section of diffractive jet production. 

For both the soft and the hard colour neutralization mechanisms the 
normalization of the jet cross sections can not be predicted from first 
principles. However, for sufficiently large transverse momenta of the 
produced jets the contribution with additional gluon and soft colour 
neutralization is expected to dominate. This is true even in the region of 
moderate diffractive masses $M$.

\end{document}